\documentclass[superscriptaddress,aps,prx,twocolumn,showpacs,nofootinbib, longbibliography, notitlepage,floatfix]{revtex4-2}
\usepackage{etex}
\usepackage{amsmath,amssymb,amsthm}
\usepackage{multirow}
\usepackage{tabularx}
\usepackage[colorlinks=true,citecolor=blue,urlcolor=blue]{hyperref}
\usepackage[pdftex]{graphicx}
\usepackage{times,txfonts}
\usepackage{braket}
\usepackage{color}
\usepackage{natbib}
\usepackage{url}
\usepackage{makecell}
\usepackage{bm}
\usepackage{amsmath,blkarray}
\usepackage{mathtools}
\usepackage{ulem}
\usepackage{latexsym}
\usepackage{tabularx, booktabs}
\usepackage{graphics,epstopdf}
\usepackage{graphicx}
\usepackage{float}
\usepackage{amsfonts}
\usepackage{tikz}
\usetikzlibrary{quantikz}
\usepackage{color,soul}

\newcommand{\be}{\begin{equation}}
\newcommand{\ee}{\end{equation}}
\newcommand{\ba}{\begin{eqnarray}}
\newcommand{\ea}{\end{eqnarray}}

\usepackage{multirow}
\usepackage{appendix}
\usepackage{url}
\usepackage{verbatim}
\renewcommand{\arraystretch}{1.7}

\begin{document}


\title{Benchmarking hybrid digitized-counterdiabatic quantum optimization}

\author{Ruoqian Xu}
\affiliation{Department of Physical Chemistry, University of the Basque Country UPV/EHU, Apartado 644, 48080 Bilbao, Spain}

\author{Jialiang Tang}
\affiliation{Department of Physical Chemistry, University of the Basque Country UPV/EHU, Apartado 644, 48080 Bilbao, Spain}

\author{Pranav Chandarana}
\affiliation{Department of Physical Chemistry, University of the Basque Country UPV/EHU, Apartado 644, 48080 Bilbao, Spain}
\affiliation{EHU Quantum Center, University of the Basque Country UPV/EHU, Barrio Sarriena, s/n, 48940 Leioa, Spain}

\author{Koushik Paul}
\email{koushikpal09@gmail.com}
\affiliation{Department of Physical Chemistry, University of the Basque Country UPV/EHU, Apartado 644, 48080 Bilbao, Spain}
\affiliation{EHU Quantum Center, University of the Basque Country UPV/EHU, Barrio Sarriena, s/n, 48940 Leioa, Spain}

\author{Xusheng Xu}
\affiliation{Department of Physics, State Key Laboratory of Low-Dimensional Quantum Physics, Tsinghua University, Beijing 100084, China}

\author{Manhong Yung}
\affiliation{Department of Physics, Southern University of Science and Technology, Shenzhen 518055, People's Republic of China}
\affiliation{Shenzhen Institute for Quantum Science and Engineering, Southern University of Science and Technology, Shenzhen 518055, People's Republic of China}
\affiliation{Guangdong Provincial Key Laboratory of Quantum Science and Engineering, Southern University of Science and Technology, Shenzhen 518055, People's Republic of China}
\affiliation{Shenzhen Key Laboratory of Quantum Science and Engineering, Southern University of Science and Technology, Shenzhen 518055, People's Republic of China}

\author{Xi Chen}
\email{chenxi1979cn@gmail.com}
\affiliation{Department of Physical Chemistry, University of the Basque Country UPV/EHU, Apartado 644, 48080 Bilbao, Spain}
\affiliation{EHU Quantum Center, University of the Basque Country UPV/EHU, Barrio Sarriena, s/n, 48940 Leioa, Spain}




\begin{abstract}
Hybrid digitized-counterdiabatic quantum computing (DCQC) is a promising approach for leveraging the capabilities of near-term quantum computers, utilizing parameterized quantum circuits designed with counterdiabatic protocols. However, the classical aspect of this approach has received limited attention. In this study, we systematically analyze the convergence behavior and solution quality of various classical optimizers when used in conjunction with the digitized-counterdiabatic approach. We demonstrate the effectiveness of this hybrid algorithm by comparing its performance to the traditional QAOA on systems containing up to 28 qubits. Furthermore, we employ principal component analysis to investigate the cost landscape and explore the crucial influence of parameterization on the performance of the counterdiabatic ansatz. Our findings indicate that fewer iterations are required when local cost landscape minima are present, and the SPSA-based BFGS optimizer emerges as a standout choice for the hybrid DCQC paradigm.
\end{abstract}

\flushbottom
\maketitle
\thispagestyle{empty}

\section{Introduction}
\label{Sec1:intro}
In the field of quantum computing, one crucial aspect of research involves benchmarking quantum algorithms. The purpose of benchmarking is to thoroughly evaluate the performance of a quantum algorithm using a well-defined set of metrics. This evaluation process is particularly important for the Noisy Intermediate-Scale Quantum (NISQ) era. Benchmarking quantum algorithms is essential for gaining insights into the capabilities and limitations of NISQ computers while executing quantum algorithms.

In recent years, the field of quantum computing has witnessed remarkable progress, opening up new frontiers for solving complex computational problems that were once considered intractable for classical computers. Among several promising quantum algorithms, Variational Quantum Algorithms (VQAs) have emerged as a particularly intriguing and versatile approach \cite{Cerezo2021}. VQAs utilize the unique properties of quantum systems to tackle optimization tasks by iteratively optimizing the parameters of a quantum circuit to minimize a given cost function. This flexibility and adaptability make VQAs well-suited for a wide range of applications, including quantum chemistry simulations~\cite{PhysRevX.8.011021,Adaptvqe,Kandala2017}, machine learning \cite{PhysRevX.11.031070,PhysRevA.108.042611}, and combinatorial optimization ~\cite{10.1007/978-3-030-14082-3_7,Karamlou2021}.

One powerful example of VQA is the Quantum Approximate Optimization Algorithm (QAOA) \cite{farhi2014quantum}. QAOA employs $p$ layers of unitary operations to iteratively explore the potential solutions using classical optimization and enhance the probability of obtaining the optimal solution. By adjusting the number of layers, QAOA offers efficient use of quantum hardware for solving a given optimization problem. One of the main challenges of QAOA is its sensitivity to the choice of hyperparameters. Finding the optimal values for these parameters can be a challenging and time-consuming task, especially for large-scale optimization problems, as it may require significant classical computational resources and extensive experimentation.

Over the past few years, numerous algorithms have been introduced to enhance the capabilities of original QAOA. One such example is multi-angle QAOA (maQAOA), which incorporates additional parameters into the quantum circuit to achieve improved approximation ratios compared to the standard QAOA \cite{herrman_multi-angle_2022}. Similarly, Adaptive Derivative-Assembled Problem-Tailored QAOA (ADAPT-QAOA) employs an iterative process to select the ansatz from a predefined pool of operators \cite{PhysRevResearch.4.033029}. This selection aims to maximize the gradient of the commutator between the pool operator and the cost Hamiltonian, leading to enhanced optimization outcomes. Several other techniques, such as Recursive QAOA (RQAOA) \cite{RQAOA1, RQAOA2}, Quantum Alternating Operator Ansatzes (QAOAnsatz) \cite{a12020034}, Spanning Tree QAOA (ST-QAOA) \cite{9585416}, and Adaptive Bias QAOA (ab-QAOA) \cite{PhysRevResearch.4.023249}, have been proposed with the aim of enhancing various aspects of the QAOA. These methods claim varying degrees of improvements in circuit depth, parameter space, operator pool, and computational cost \cite{lockwood2022empirical}.
\begin{table*}[t]
\centering
\caption{Update rules of parameters and gradients for hybrid gradient-based-optimizers using parameter shift rules (PS) or SPSA rules and gradient-free-optimizers.}
\begin{ruledtabular}
\tabcolsep=1cm
\renewcommand\arraystretch{2}
\begin{tabular}{ccc}
  Optimizers & Gradient calculation & Update rules \\ \hline
    \makecell*[c]{SPSA-Adam\\SPSA-SGD\\SPSA-BFGS}& \makecell*[c]{$\nabla_{\bm{\theta}} J(\bm{\theta}) = \frac{J(\bm{\theta}+c\bm{\delta})-J(\bm{\theta}-c\bm{\delta})}{2c\bm{\delta}}$} & \makecell*[c]{$\bm{\theta}_{k+1} = \bm{\theta}_{k} - \frac{\eta}{\sqrt{\hat{v_{k}}}+\epsilon}\hat{m_{k}}$\\$\bm{\theta}_{k+1} = \bm{\theta}_{k} - a_{k}\cdot \nabla_{k}J(\bm{\theta})$\\$\nabla_{k}J(\bm{\theta}) = \nabla_{k
    +1}J(\bm{\theta}^{k+1}) + B_{k+1}(\bm{\theta}^{k}-\bm{\theta}^{k+1})$} \\
  \hline
  \makecell*[c]{PS-Adam\\PS-SGD\\PS-BFGS}& \makecell*[c]{$\nabla_{\theta_{i}}J(\bm{\theta}) = r[J(\theta_{i} + \frac{\pi}{4r}) -J(\theta_{i} -\frac{\pi}{4r})]$} & \makecell*[c]{$\bm{\theta}_{k+1} = \bm{\theta}_{k} - \frac{\eta}{\sqrt{\hat{v_{k}}}+\epsilon}\hat{m_{k}}$\\$\bm{\theta}_{k+1} = \bm{\theta}_{k} - a_{k}\cdot \nabla_{k}J(\bm{\theta})$\\$\nabla_{k}J(\bm{\theta}) = \nabla_{k
    +1}J(\bm{\theta}^{k+1}) + B_{k+1}(\bm{\theta}^{k}-\bm{\theta}^{k+1})$} \\[1em] \hline
  COBYLA  & \multicolumn{2}{c}{Terminates when the target radius is satisfied.}                         \\ \hline
  Nelder-Mead  & \multicolumn{2}{c}{Terminates when the cost function is minimized according to the criteria in Eq.~\ref{eq:nm}.} 
  \\ 
\end{tabular}
\end{ruledtabular}
\label{tab:8op}
\end{table*}

In this paper, we examine the performance and effectiveness of hybrid digitized-counterdiabatic quantum computing (DCQC) algorithms \cite{PhysRevResearch.4.013141}. This method incorporates elements from Shortcut to Adiabaticity (STA) \cite{insta9,sta16}, particularly the use of counterdiabatic (CD) driving \cite{PhysRevLett.111.100502}, to reduce the circuit depth and improve the optimization process of conventional QAOA. In prior studies, these algorithms have exhibited their efficacy in addressing diverse optimization problems, including MaxCut, portfolio optimization, protein folding, and several quantum many-body systems \cite{PhysRevResearch.4.013141,nar,PhysRevA.104.L050403,PhysRevResearch.4.043204,chandarana2023digitized}. We examine the classical component of the hybrid DCQC in this study. Our investigation focuses on convergence while using different gradient-based and gradient-free optimizers. 
We illustrate the scalability of this approach concerning system size, highlighting its effectiveness in finding the ground state, even when 
$p = 1$  within systems containing up to 28 qubits. To consolidate these findings, we explore the Fourier landscape of the cost function for various parameterizations using principal component analysis (PCA).

This paper is structured as follows. Section.~\ref{sec2:prelims} provides a brief discussion of hybrid DCQC followed by a detailed description of gradient-based and gradient classical optimizers and their application in hybrid DCQC. This is followed by respective findings regarding various parameterizations, cost landscape, and variance of gradients for different optimizers in Section.~\ref{sec3:results}. We finally conclude in Section.~\ref{sec4:conclusion} and provide direction for future work.

\section{Background}
\label{sec2:prelims}
Hybrid DCQC algorithms fall under the umbrella of VQAs where the quantum part consists of a circuit ansatz with parameterized gates. These parameters are optimized by classical routines to minimize a given cost function. The selection of these parameterized gates can be done in two major ways: taking information from the problem or taking information from the hardware. The former is known as problem-inspired ansatz and the latter is known as hardware-efficient ansatz. One of the popular problem-inspired ansatzes is QAOA where two unitaries $U_b(\bm{\beta})$ and $U_c(\bm{\alpha})$ with tunable parameters $( \bm{\alpha},\bm{\beta})$ are applied iteratively $p$ times to an initial state $\ket{\psi_0}$. Here, $U_b(\bm{\beta}) = \exp{(-i \bm{\beta} H_m})$ is the mixing unitary and $U_c(\bm{\alpha}) = \exp{(-i \bm{\alpha} H_c})$ is the unitary corresponding to the cost Hamiltonian with $[H_m, H_c] \neq 0$. The aim is to minimize the expectation value $\braket{\psi(\bm{\alpha},\bm{\beta})|H_c|\psi(\bm{\alpha},\bm{\beta})}$ for achieving the ground state of the system that encodes the solution the chosen problem.

The main drawback of QAOA is the need for high circuit depth (large $p$) as it takes inspiration from the adiabatic process which is inherently slow. To circumvent this, DCQC makes use of counterdiabatic protocols to decrease the circuit depth of the ansatz. In DCQC, the circuit ansatz is chosen by finding suitable operators using the nested commutator (NC) method~\cite{PhysRevLett.123.090602}. The resultant parameterized unitary is of the form~\cite{chandarana2023digitized}
\begin{equation}
  U_d(\bm{\theta})=\exp{ \big( -i \bm{\theta} \mathcal{A} \big)},   
\end{equation}
where $\mathcal{A} \in A^{(l)}_\lambda$, and $A^{(l)}_\lambda$ is an operator pool obtained from the NC method \cite{chandarana2023digitized}. $\bm{\theta}$ is the set of parameters to be optimized. 

The classical optimizers, employed to optimize these parameters, also play a crucial role in deciding the performance of an ansatz. Classical optimizers can be broadly classified into two categories: gradient-based and gradient-free. In gradient-based optimization, the iteration step is decided based on the gradient direction. The gradient values can be evaluated by various methods such as finite-difference or parameter-shift~\cite{PhysRevA.99.032331}. There are also newer methods that take fewer circuit evaluations to compute the gradient~\cite{hoffmann2022gradient}. After evaluating the gradient, classical optimizers like Adam~\cite{kingma2014adam} and Adagrad~\cite{JMLR:v12:duchi11a} are implemented to find the next iteration steps. 
 
On the other hand, gradient-free optimizers do not rely on the gradient information to determine the next iteration. These include optimizers like COBYLA~\cite{Powell1994}, Nelder-Mead~\cite{10.1093/comjnl/7.4.308}, etc. There also exist other optimizers like the simultaneous perturbation stochastic approximation (SPSA)~\cite{119632} that do not directly utilize gradient information. In SPSA, the gradient is approximated along a randomly chosen direction by a single partial derivative computed by finite difference. This has shown a faster convergence rate than the finite difference method~\cite{119632}.

Before starting to explore gradient-based optimizers, we commence with describing a few methods to calculate gradients which will be used later.

\textit{Parameter shift (PS):}~Let us consider a cost function $J(\bm{\theta})$, which is parameterized by some variables $\bm{\theta}\in \mathbb{R}$. PS enables us to calculate the exact gradient of $J(\bm{\theta})$ with respect to {$\bm{\theta} = \{\theta_{1},~\theta_{2},~\cdots,~\theta_{m}\}$} using the following relation \cite{PhysRevA.99.032331}:
\begin{equation}
\nabla_{\theta_{i}}J(\bm{\theta}) = r[J(\theta_{i} + \frac{\pi}{4r}) -J(\theta_{i} -\frac{\pi}{4r})],
\end{equation}
where, $\nabla_{\theta_{i}}J(\bm{\theta})$ represents the gradient with respect to element $\theta_{i} \in  \bm{\theta}$. $r$ is the shift constant, determined by the ansatz chosen. For instance, it is proved that $r = \frac{1}{2}$ for all single qubit gates \cite{PhysRevA.98.032309}. This method evaluates the same ansatz twice {based on different shifted parameters} to find the gradient making the runtime proportional to twice the total number of parameters.

\textit{SPSA:}~ SPSA approaches the gradient by introducing a $m$-dimensional perturbation vector $\bm \delta =(\delta_{1},\delta_{2},\cdots,\delta_{m})^{T}$, whose elements $ \delta
_{i}$ are all randomly picked from $\{-1,1\}$. The gradient estimator can be written as \cite{705889}, 
\begin{equation}\label{eq:spsa}
\nabla_{\bm{\theta}} J(\bm{\theta}) = \frac{J(\bm{\theta}+c\bm{\delta})-J(\bm{\theta}-c\bm{\delta})}{2c\bm{\delta}},
\end{equation}
where $c$ is a small positive scaling factor. Due to this perturbation, SPSA is robust under noisy environments.

\textit{Adjoint differentiation:}
Consider the circuit ansatz defined by the unitary evolution operator $U =$$ U_{m}(\theta_{m})\cdots U_2(\theta_2)U_1(\theta_1)$ and any observable $\mathcal{\hat{O}}$ for which we compute the trial state as {$\ket{\varphi_{i}}=U_{i+1}^{\dagger}(\theta_{i+1})\cdots U_{m}^{\dagger}(\theta_{m})\mathcal{\hat{O}}U_{m}(\theta_{m})\cdots U_2(\theta_2)U_1(\theta_1)\ket{0}$ and define $\ket{\psi_{i}}=U_{i}(\theta_{i})\cdots U_2(\theta_2)U_1(\theta_1)\ket{0}$, such that the gradient with respect to $\theta_{i}\in \bm{\theta}$ can be formulated as,
\begin{equation}\label{eq:adjoint}
     \nabla_{\theta_{i}}J(\bm{\theta}) = \frac{\partial \bra{0}U^\dagger \mathcal{\hat{O}} U \ket{0}}{\partial \theta_i} =2\textbf{Re} \bra{\varphi_{i-1}}\frac{\partial U_i(\theta_i)}{\partial \theta_i}\ket{\psi_{i-1}}.
\end{equation}}

\begin{figure}
    \centering
    \includegraphics[width=1\linewidth]{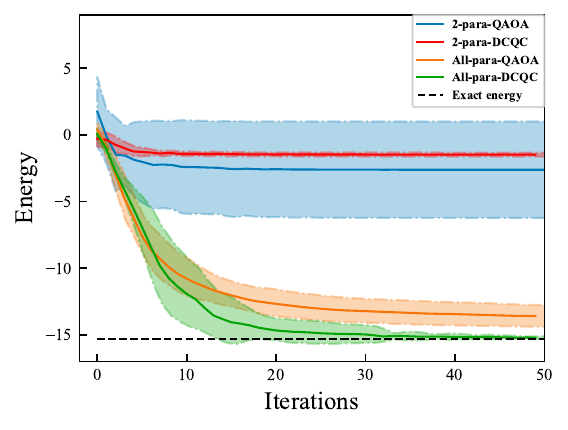}
    \caption{Energy as a function of iterations for the $p = 1$ layer, $N = 10$ qubits. Results show simulator data for $50$ iterations using the PS-Adam optimizer comparing QAOA and DCQC ansatz. Ten initial parameters are randomly chosen for each ansatz. The shaded area shows the standard deviation.}
    \label{fig:4 comparison}
\end{figure}

Using the above-mentioned methods of computing gradients, we implement a total of six gradient-based optimizers, which are PS-SGD, PS-BFGS, PS-Adam, SPSA-SGD, SPSA-BFGS, SPSA-Adam, to compare their performance with different ansatzes. {Further details on the optimizers can be found in Appendix 
\ref{app:gb}.} In addition, we also used two gradient-free optimizers, namely, COBYLA, and Nelder-Mead (See Appendix \ref{app:gf}). Table \ref{tab:8op} provides a brief summary of the update rules for every optimizer.
{Due to the high computational costs required for parameter shift, we use the adjoint method in simulations. This is because we have found a strong similarity between these two methods of computing gradients, which is explained explictly in Appendix~\ref{app:ps}.} 

One of the most important factors that {determines} the performance of an ansatz is its expressibility, which is defined as how uniformly the ansatz is capable of searching the solution space of the problem Hamiltonian. When the solution is not known, a highly expressible ansatz is required to explore larger spaces such that the probability of finding the solution is higher. On the other hand, even when the ansatz includes the solution, it should be trainable, such that the optimization can reach the solution effectively. So, the requirement for a 'good' ansatz is that it should include the solution and be {sufficiently} trainable. Therefore, the choice of classical optimization can have a drastic impact on the trainability of the circuit ansatz. High expressibility is not always as good as one may face the problem of barren plateaus where the gradient vanishes exponentially with increasing system size and parameter space. Random circuits have already shown the presence of barren plateaus with increasing system size~\cite{mcclean2018barren}. 

Another crucial element related to expressibility and trainability is the parameterization of the ansatz. Several works have been done to investigate the effect of parameterization on the performance of the ansatz chosen~\cite{Larocca2023}. Thus, making the right choice of parameterization and classical optimizers is crucial for an optimization algorithm to work. For example, if we have many local minima in the energy landscape, even with a good classical optimizer, we can get stuck to solutions that are not close to what we expect. Therefore, analysis of the energy landscape becomes important. If parameters scale with the system size, the energy landscape becomes high dimensional which makes the analysis non-trivial. 

PCA is a commonly utilized technique in data
analysis, aimed at reducing the dimensionality of a dataset while preserving as much information as possible. Generally, this is done by transforming the dataset to a new space determined by principal components, through which a lower dimensional hyperplane can be obtained such that the visualization of multidimensional data becomes easier. The computation of the principal components is done using the eigenvectors of the covariance matrix generated by the dataset. 

Here, we employ PCA to reduce the dimensionality of the $m$-dimentional parameter space generated by DCQC ansatz, $\bm{\theta} = (\theta_1, \theta_2, \cdots, \theta_m)^T$. Using PCA, we identify the $n$ ($n \leq m$) number of principal components which represent the highest variance among the parameters during the optimization process. The principal components are determined by the orthogonal eigenvectors $\vec{\mathcal{E}}_i$ of the covariance matrix $\mathcal{E}$ where the eigenvector corresponding to the largest eigenvalue gives the direction of maximum variance. The original dataset and the principal component space are related by this unitary projection matrix $\mathcal{E}$ such that,
\begin{equation}
\bm{\gamma} = \mathcal{E}\cdot \bm{\theta},
\label{eq:PCA}
\end{equation}
where $\bm{\gamma} = ({\gamma}_1, {\gamma}_2, \cdots, {\gamma}_m)$ represents the principal components and $\mathcal{E} = (\vec{\mathcal{E}}_1,  \vec{\mathcal{E}}_2, \cdots, \vec{\mathcal{E}}_m)^T$ defines its direction. Each $\gamma_{i}$ contains the information about the variance of $i$-th principal component. A reduced $n$-dimensinal principal component space can thus be achieved by truncating at $n$ position. To obtain a more profound comprehension of optimization process using different optimizers for hybrid DCQC, we implement the first two principal components to demonstrate the cost landscape. Corresponding original parameters $\bm {\theta}$ can then be obtained using Eq.~\ref{eq:PCA} and utilized to visualize the optimization trajectory on the cost landscape.

\begin{figure}
    \centering
    \includegraphics[width=1\linewidth]{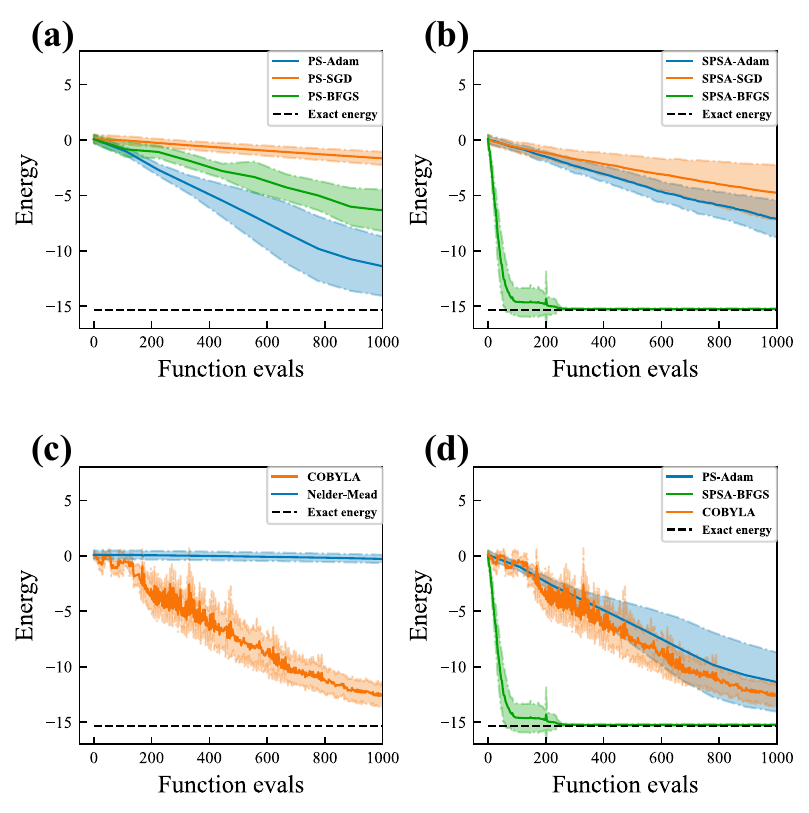}
    \caption{Energy as a function of a number of function evaluations for (a) PS-based optimizers (b) SPS-based optimizers (c) gradient-free optimizers (d) the best ones from the previous group. Comparisons are made between different optimizers on $p = 1$, $N = 10$ qubits system using fully parameterized DCQC ansatz. Ten random initial parameters are chosen to be the same for each optimizer.}
    \label{fig:gb-gf}
\end{figure}
\section{Results}
\label{sec3:results}
In this section, we benchmark hybrid DCQC ansatz with different parameterizations as well as different optimizers to find the ground state of the Sherrington-Kirkpatrick (SK) model. SK model is a classical spin model that possesses all-to-all connectivity~\cite{RevModPhys.58.801,PhysRevLett.35.1792}. This model is particularly interesting because several optimization problems can be encoded into spin glass systems~\cite{10.3389/fphy.2014.00005}. The Hamiltonian of the SK model is given by
\begin{equation}
	H_c=-\sum_{i, j } J_{i j} \sigma_{z}^{i} \sigma_{z}^{j},
\end{equation}
where $J_{ij}$ are coupling coefficients associated to spin $i$ and spin $j$. We study the SK model with qubits ranging from $N=14$ to $N=28$ qubits with $J_{ij} \in \{-1,1\}$ with mean $0$ and variance $1$. For this model, the favourable CD ansatz is $\mathcal{A} = \sum_{i }\sigma^{i}_{y} + \sum_{i, j }\sigma^{i}_{y}\sigma^{j}_{z} $.   

In our investigation of the parameterization, we consider two types of ansatz: a 2-parameter ansatz, where one parameter is associated with all the single-qubit gates, and the other with the two-qubit gates. The second one is a fully parameterized ansatz, wherein each gate contains its own independent free parameter which amounts to a total of $N(N-1)/2 + N$ number of free parameters. In Fig.~\ref{fig:4 comparison}, we study the impact of this parameterization in both conventional QAOA and hybrid DCQC for $p = 1$, utilizing the PS-ADAM optimizer for $N=10$ qubits. The $2$-parameter ansatz fails to reach the ground state where QAOA shows better minimization compared to the hybrid DCQC ansatz. However, DCQC shows a lower standard deviation (the shaded region) which means it quickly converges to particular minima. On the other hand, QAOA shows high standard deviation showing random behaviour in the convergence process. Upon fully parameterizing the ansatz, DCQC shows superior performance in efficiently reaching the ground state.  The standard deviation is also small for the final solution in DCQC proving its efficacy against gradient-based optimizers. Note that this fully parameterized version of QAOA is known as maQAOA~\cite{herrman_multi-angle_2022} which has been shown to have a superior performance than QAOA. That being said, increasing the number of parameters per layer will result in an increase in expressibility. From this, we can conclude that the fully parameterized DCQC ansatz performs the best. Therefore, we select this ansatz and perform further analysis with various optimizers. 
\begin{figure}
    \centering
    \includegraphics[width=1\linewidth]{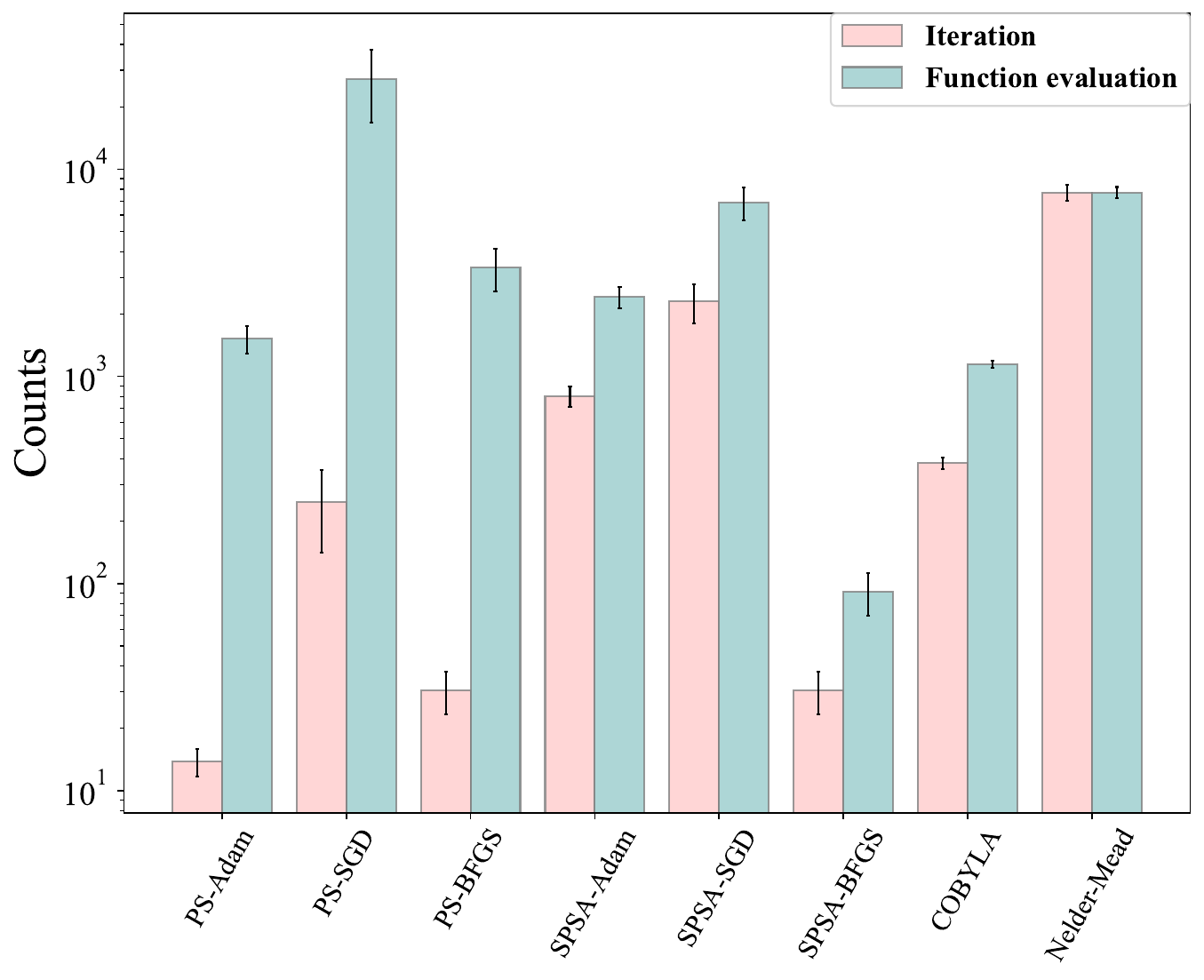}
    \caption{Number of operations as a function of optimizer type for 10 instances of SK model when $p = 1$ layer, $N = 10$ qubits. The figure shows the counts (iterations, function evaluations) needed to reach the fidelity of 90\% using SPSA-BFGS with fully parameterized DC ansatz. The gray lines at the tips of the bars represent the width of each standard error.}
    \label{fig:ite-fev}
\end{figure}

To compare the impact of different classical optimizers, we restrict ourselves to a system size of $N=10$ qubits, initialize with ten random instances. Fig.~\ref{fig:gb-gf} illustrates the performance of all the optimizers in 1000 function evaluations. The number of function evaluations is defined by the total number of measurements performed to evaluate the cost function during the optimization process. This metric is most suitable for comparing the performances of both gradient-based and gradient-free optimizers. In the case of gradient-based optimizers, this also includes the number of measurements required to compute the gradient. For instance, PS requires two function evaluations to calculate gradients with respect to each parameter resulting in $(2m+1)$ function evaluations for each iteration. Note that, gradient calculation using PS is costly in our study as the number of function evaluations increases linearly with the number of parameters. From Fig.~\ref{fig:gb-gf}a, we show the comparison of three optimizers where the gradients are computed using the PS rule. We observe that, among these PS-based optimizers, PS-Adam can approach ground-state energy faster than PS-SGD and PS-BFGS. 

In general, SPSA-based optimization tends to perform better compared to PS as it requires only three function evaluations for each iteration regardless of the size of the parameter space. For gradient-based optimizers using SPSA, we observe that BFGS outperforms the other two with significant advantages, with fewer function evaluations. To make a thorough analysis, we have also compared two gradient-free optimizers in Fig.~\ref{fig:gb-gf}c. Like SPSA, in COBYLA we also need three function evaluations for each iteration whereas  Nelder-Mead requires one evaluation only. However, COBYLA converges more efficiently compared to Nelder-Mead. In Fig.~\ref{fig:gb-gf}d, we compare the best optimizers from the three above-discussed groups where SPSA-BFGS comes out to be the best optimizer among all the optimizers studied in this paper. It is also important to note that this benchmarking heavily depends on the system size. Nonetheless, for smaller $N$, SPSA-BFGS demonstrates the best performance for DCQC. In addition, to comprehensively understand the process of reaching the minima, we count the number of iterations and the function evaluations needed for each optimizer to achieve 90\% of the exact ground state energy. Fig.~\ref{fig:ite-fev} clearly depicts the SPSA-BFGS outperforms other optimizers in terms of both the number of iterations and the function evaluations. 

In order to numerically analyze the scalability of DCQC in the NISQ regime, we conducted evaluations based on the mean of 10 randomly initialized parameters for a specific instance of the model, employing $p=1$ layers of the ansatz and the SPSA-BFGS optimizer from $N=10$ to $N=28$ qubits. 
\begin{figure}
    \centering
    \includegraphics[width=1\linewidth]{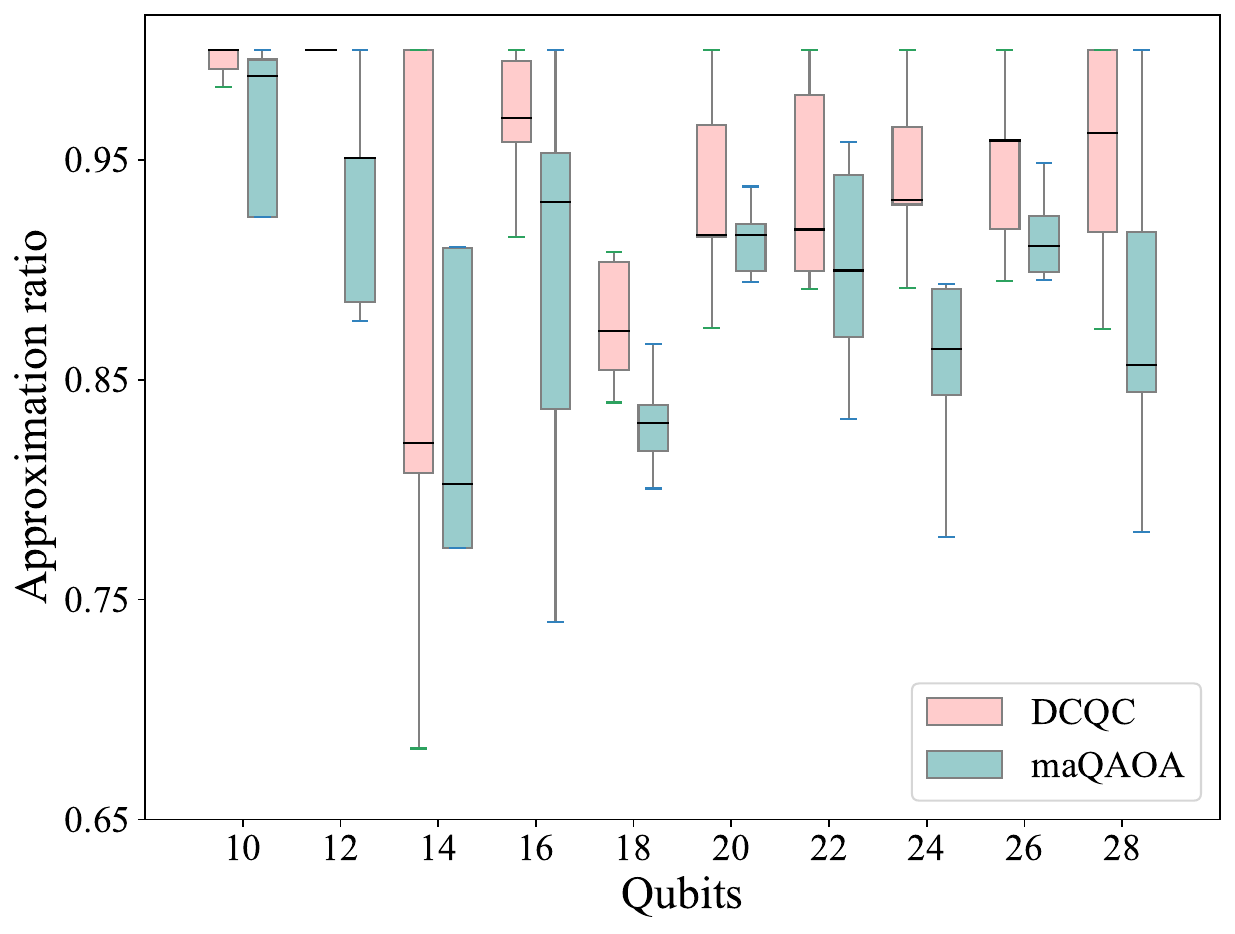}
    \caption{Approximation ratio as a function of qubit number. The result represents $p=1$ layered, $N=10$ to $N=28$ qubits ansatz. SPSA-BFGS optimizer is considered for both maQAOA and DCQC ansatz.}
    \label{fig:scaleminq}
\end{figure}
In Fig.~\ref{fig:scaleminq}, we plot the approximation ratio with respect to the number of qubits where the approximation ratio is defined as the $\braket{Hc}/E_{g}$. We observe that the performance of DCQC ansatz is better than QAOA considering the fact that the size of the parameter space is the same for both the ansatz. Fig.~\ref{fig:scaleminq} also shows that the performance of both ansatzes highly depends on the initial parameters chosen. This statement is evident from the high error bars, signifying that when the initialization is not good, the optimization lands into local minima. Nevertheless, with appropriate initial parameters, we can find the exact ground state for systems up to $N=28$. Note that even for an ansatz with $m=406$, DCQC can reach the exact ground state proving the absence of issues like barren plateau in the cost landscape.

Next, to visualise the cost landscape and the optimization trajectory, we employ PCA, which is designed to reduce the dimensionality of the parameter space by identifying a hyperplane that effectively captures the properties of all the samples. 
As mentioned in the previous section, PCA enables us to obtain the variance for each dimension of the hyperplane, which serves as a measure of the sparsity in the samples. By leveraging PCA, we gain valuable insights into the high-dimensional data, allowing us to better understand and navigate through the complex cost landscape.

\begin{figure*}
    \centering
    \includegraphics[width=1\linewidth]{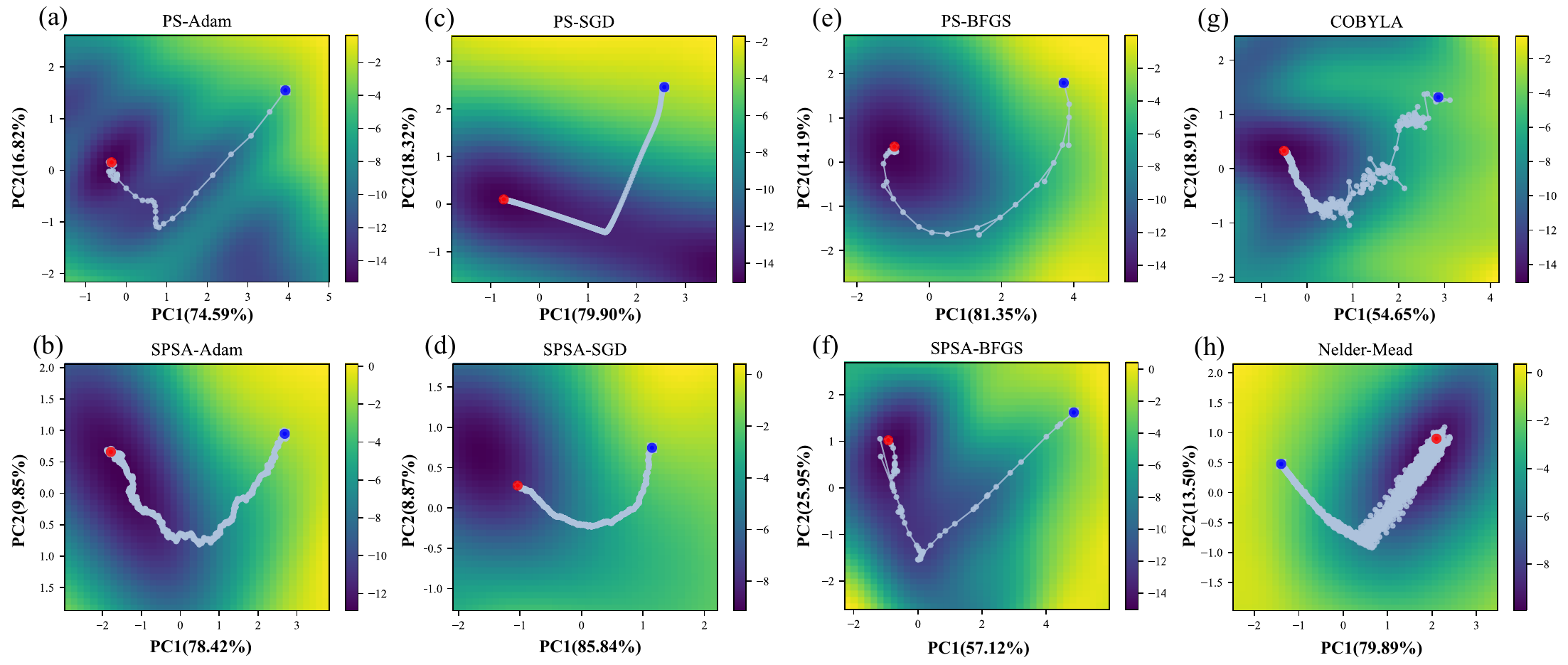}
    \caption{Principal component (PC) analysis result on the cost landscape of $p = 1$, $N = 10$ qubits system. The figure shows how the optimization trajectory goes with different hybrid optimizers on the cost landscape. The $x,y$ axis denotes how much the first and second PC will affect the variance of the data. The red and blue dots represent the final and initial points respectively.}
    \label{fig:pca-1}
\end{figure*}

The first two principal components ($x$ and $y$ axes in Fig.~\ref{fig:pca-1}) account for at least 70\% of the total variance in our data,  which permits us to effectively visualize and understand the primary trends and patterns in the cost function. This enables us to make informed decisions to refine our optimization strategies accordingly \cite{rudolph2021orqviz}.

In Fig.~\ref{fig:pca-1}, we start the optimization with the same initial points for all the optimizers (shown by the blue dots) and show the optimization trajectory that eventually reaches to the minimum (the red dots). The number of points represents the number of iterations required to obtain the global minimum. The observations of the cost landscape validate the results we obtained in Fig.~\ref{fig:gb-gf}. We can observe that, the optimization requires less number of iterations when local minima exist in the cost landscape of the principal components. As depicted in Fig.~\ref{fig:pca-1}a and Fig.~\ref{fig:pca-1}f, ADAM works best for the PS-based optimization whereas, among SPSA based optimizers, BFGS reaches the minima in the fewest possible steps. Furthermore, cost landscapes in Fig.~\ref{fig:pca-1}d and Fig.~\ref{fig:pca-1}h indicate the inability of SPSA-SGD and nelder-Mead to reach a global optimum. 

Considering the results presented earlier, we can confidently assert that SPSA based BFGS optimizer emerges as the most promising candidate for hybrid DCQC. Its effectiveness, combined with its ability to efficiently navigate through the cost landscape, makes it most suitable for the optimization process. This conclusion underscores the importance of selecting appropriate classical optimizers to enhance the performance and practicality of the DCQC algorithm for various quantum computing applications.

\section{Conclusion}
\label{sec4:conclusion}

In this article, we have conducted an extensive study of a hybrid DCQC algorithm, bench-marking its performance and efficacy with respect to various classical optimizers. Our results suggest that gradient-based optimizers like ADAM and BFGS typically exhibit faster convergence in comparison to gradient-free optimizers such as COBYLA and Nelder-Mead as they need less function evaluations. Nonetheless, the method used for gradient calculation plays a pivotal role in distinguishing the most efficient optimizer. When the number of circuit evaluations is considered as the performance metric, SPSA-BFGS emerges as the best, demonstrating a distinct advantage over other optimizers, whether they are gradient-based or gradient-free. For a fully parameterized ansatz, DCQC with SPSA-BFGS provides a better approximation ratio compared to maQAOA even when it is scaled up to a system size $N=28$, avoiding the occurrence of any trainability issues such as barren plateaus. Moreover, the study of the cost landscape yields valuable insights into the behavior of the optimization processes, particularly in relation to variations in the first two principal components obtained through PCA. However, one issue requires addressing. The assumption of equality between parameter shift and adjoint derivative is founded solely on analytic, and further validation is necessary as the library continues to be developed. 

In summary, this study serves as a comprehensive benchmark for hybrid DCQC algorithms, specifically examining their performance in optimizing the parameter space and its correlation with the cost landscape. Through an analysis of various gradient-based and gradient-free optimizers, valuable insights were gained into their convergence rates, efficiency, and sensitivity to system size and initial parameters. Additionally, employing PCA to study the cost landscape shed light on the optimization trajectory, deepening our understanding of the process itself. Notably, our emphasis on scalability sets this research apart. By evaluating DCQC in a system with up to 28 qubits, we underscore the substantial potential and relevance of this approach in addressing larger-scale quantum computing challenges. 
This research establishes a basis for selecting classical optimizers in hybrid DCQC techniques, enhancing their potential applications across diverse domains to address complex problems.

\begin{acknowledgments}

This work is supported by the Basque Government through Grant No. IT1470-22, the project grant PID2021-126273NB-I00, PID2021-123131NA-I00 funded by MCIN/AEI/10.13039/501100011033 and by ``ERDF A way of making Europe" and ``ERDF Invest in your Future", NSFC (12075145), and STCSM (2019SHZDZX01-ZX04), the China Scholarship Council (CSC) under Grant Nos: 202206890003, 202306890004. This project has also received support from the Spanish Ministry of Economic Affairs and Digital Transformation through the
QUANTUM ENIA project call-Quantum Spain, and by the
EU through the Recovery, Transformation and Resilience
Plan–NextGenerationEU within the framework of the Digital
Spain 2026 Agenda. X.C. acknowledges ayudas para contratos Ram\'on y Cajal–2015-2020 (RYC-2017-22482). 

\end{acknowledgments}

\begin{appendices}
\appendix

\section{Gradient-based Optimizers}\label{app:gb}
The optimizer processes the training by adjusting the model's parameters iteratively based on the gradients of the cost function with respect to these parameters. The primary goal is to find the optimal set of parameters that lead to the best performance of the model on the given task \cite{wiedmann2023empirical}. Three gradient-based optimizers will be introduced in this section.\\

\textit{Stochastic gradient descent (SGD)}~
Normally, its advantage lies in using small batches in each iteration rather than entire data sets. In our case, the parameter is fed one by one and updated according to this rule: 
\begin{equation}
    \bm{\theta}_{k+1} = \bm{\theta}_{k} - a_{k}\cdot \nabla_{k}J(\bm{\theta}),
\end{equation}
where $a_{k} = \frac{a}{(A+k)^{b}},~b,~a,~A$ are all scaling parameters to control the learning rate and $k$ is the iteration number. 

\textit{Broyden–Fletcher–Goldfarb–Shanno (BFGS)}~
The BFGS algorithm belongs to a class of quasi-Newton optimization methods that extend Newton's method \cite{nocedal2006numerical}. 
\begin{equation}
\nabla_{k}J(\bm{\theta}) = \nabla_{k
    +1}J(\bm{\theta}^{k+1}) + B_{k+1}(\bm{\theta}^{k}-\bm{\theta}^{k+1}),
\end{equation}
where $B_{k+1}\Delta{\theta^{k}}= y^{k}$ is
defined.
The algorithm's update formula is given by imposing constraints on the inverse Hessian B, we can produce the update following this rule: 
\begin{equation}
    B_{k+1} = B_{k} + \frac{y^{k}(y^{k})^{T}}{(y^{k})^{T}\Delta\bm{\theta}^{k}}-\frac{B_{k}\Delta{\bm{\theta}^{k}}(\Delta{\bm{\theta}^{k}})^{T}B_{k}}{(\Delta\bm{\theta}^{k})^{T}B_k\Delta{\bm{\theta}^{k}}},
\end{equation}
The only requirement for updating the value is to know the previous gradient information. 

\textit{Adaptive Moment Estimation (Adam)}~
To work with the Adam optimizer, we define the first moment (the mean) $m$ and the second moment (the variance) $v$ of each gradient respectively \cite{chandarana2022meta}. 
\begin{align}
    m_{k} &= a_{1}m_{k-1} + (1-a_{1})g_{k},\\
    v_{k} &= a_{2}v_{k-1} + (1-a_{2})g_{k}^{2},
\end{align}
with $a_{1},~a_{2}$ being the decay rates. Therefore, we have the momentums for each iteration 
\begin{align}
    \hat{m_{k}} = \frac{m_{k}}{1-a_{1}^{k}},\\
    \hat{v_{k}} = \frac{v_{k}}{1-a_{2}^{k}},
\end{align}
With these, we can write the updated formula for Adam:
\begin{equation}\label{eq:adam}
    \bm{\theta}_{k+1} = \bm{\theta}_{k} - \frac{\eta}{\sqrt{\hat{v_{k}}}+\epsilon}\hat{m_{k}},
\end{equation}
where default values are $a_{1}=0.9,~a_{2}=0.999,~\epsilon=10^{-8}$ in the neural network module of Mindquantum \cite{mq_2021}. \\

\section{Gradient-free Optimizers}\label{app:gf}
In the following sections, we will briefly introduce two methods that don't require calculating gradients.\\

\textit{Constrained optimization by linear approximation (COBYLA)}\quad
This is a model-based local optimization method that builds a linear approximation of the objective function over a simplex.
We use it through SciPy, which follows the principles below \cite{powell_1998}:
\begin{itemize}
    \item[1.] Start from $\rho = \rho_{i}, \bm{\theta_{0}} = \bm{\theta_{i}}$, the optimizer evaluates $J(\bm{\theta})$ then from an $\bm{\theta}$ such that $\bm{\theta} - \bm{\theta}_{i} \leq \rho$, calculates $J(\bm{\theta}_{i})$.
    \item[2.] Compares its value with $J(\bm{\theta})$. If the new value $\bm{\theta}_{i} $ gives $J(\bm{\theta}_{i}) \leq J(\bm{\theta})$ then $\bm{\theta} = \bm{\theta}_{i}$ is the new starting point for the next iteration.
    \item[3.] The control distance $\rho$ is decreased during iterations in order to evaluate the minimum of the function as accurately as possible.
    \item[4.] The optimization algorithm finishes when the control radius $\rho$ becomes less than a predefined value $\rho_{end}$ (fixed to $10^{-5}$).
\end{itemize}
In other words, the initial input of the algorithm is a starting point $\theta_{0}$, a control radius initial $\rho_{i}$, and a target radius $\rho_{end}$ \cite{inproceedings}.

\textit{Nelder-Mead}\quad
For the Nelder-Mead optimizer, it compares the values of $J(\theta_{i})$ at the n iteration and a simplex, trying to approximate the object function with the simplex. A simplex with size a is initialized at $\theta_{0}$ and we have \cite{LUERSEN20042251}
\begin{equation}
    \bm{\theta}_{i} = \bm{\theta}_{0} + p\vec{e}_{i} + \sum_{k=1, k\neq{i}}^{n}q\vec{e}_{k},
\end{equation}
where $\vec{e}_{i},~\vec{e}_{k}$ are unit base vectors and are defined by these
\begin{align}
    p &= \frac{a}{n\sqrt{2}}(\sqrt{n+1}+n-1),\\
    q &= \frac{a}{n\sqrt{2}}(\sqrt{n+1}-1),
\end{align}
The algorithm carries out reflection, expansion, and contraction operations to update the cost function, with the aim of similizing the function value \cite{nocedal2006numerical}. If the algorithm can terminate depends on the following criteria:
\begin{equation}\label{eq:nm}
    \sqrt{\sum_{i=1}^{n+1}\frac{(J(\bm{\theta}_{i})-\overline {J(\bm{\theta})})^{2}}{n}} < \epsilon, \overline {J(\bm{\theta})} = \frac{1}{n+1}\sum_{i=1}^{n+1}J(\bm{\theta}_{i}),
\end{equation}
where $\epsilon$ is a infinitesimal positive scalar.  

\section{Parameter shift and adjoint differentiation}\label{app:ps}
Given a quantum circuit with $\displaystyle m$ unitary operators as $\displaystyle \{U_{0}( \theta _{0}) ,...,U_{i}( \theta _{i}) ,...,U_{m}( \theta _{m}) \}$, we calculate the expectation of an observable $\displaystyle \hat{O}$:
\begin{equation}
J( \theta ) = {\bra{0}}\prod_{i=0}^{m} U_{i}^{\dagger }( \theta _{i}) \hat{O} \prod_{j=m}^{0} U_{j}( \theta _{j})\ket{0}.
\end{equation}
We set the notions like this: 
\begin{align}
    \ket{\psi _{i}} &=\prod_{j=i}^{0}U_{j}( \theta _{j})\ket{0},\\
    B_{i} &=\prod_{j=i}^{m}U_{j}^{\dagger }( \theta _{j}) \hat{O} {\prod_{k=m}^{i}} U_{k}( \theta _{k}).
\end{align}
So that we can calculate the gradient of $J( \bm{\theta} )$ with respect to parameter $\theta _{i}$:
\begin{equation}
\frac{\partial J( \bm{\theta} )}{\partial \theta _{i}} =\frac{\partial }{\partial \theta _{i}}\bra{\psi _{i-1}} U_{i}^{\dagger }( \theta _{i}) B_{i+1} U_{i}( \theta _{i})\ket{\psi _{i-1}}.
\end{equation}

Using adjoint differentiation method to calculate the gradient:
\begin{align}
\frac{\partial J( \bm{\theta} )}{\partial \theta _{i}} & =\bra{\psi _{i-1}}\frac{\partial U_{i}^{\dagger }( \theta _{i})}{\partial \theta _{i}} B_{i+1} U_{i}( \theta _{i})\ket{\psi _{i-1}}\notag\\
 & +\bra{\psi _{i-1}} U_{i}^{\dagger }( \theta _{i}) B_{i+1}\frac{\partial U_{i}( \theta _{i})}{\partial \theta _{i}}\ket{\psi _{i-1}}\notag\\
 & =2\text{Re}\left(\bra{\psi _{i-1}} U_{i}^{\dagger }( \theta _{i}) B_{i+1}\frac{\partial U_{i}( \theta _{i})}{\partial \theta _{i}}\ket{\psi _{i-1}}\right)\notag\\
 & =2\text{Re}\left(\bra{\varphi _{i-1}}\frac{\partial U_{i}( \theta _{i})}{\partial \theta _{i}}\ket{\psi _{i-1}}\right),
\end{align}
where $\displaystyle \bra{\varphi _{i-1}} =\bra{\psi _{i-1}} U_{i}^{\dagger }( \theta _{i}) B_{i+1}$ is set for simplification, so that it is easy to calculate $\displaystyle \partial J(\bm{\theta})/\partial \theta _{i-1}$.
\newline

\indent Using parameter shift method to calculate gradient, now we support $\displaystyle U( \bm{\theta} ) =e^{-i\theta \hat{P}/2}$, where $\displaystyle \hat{P}=\otimes _{i=0}^{n} \sigma _{i},~\sigma _{i} \in \{I,~X,~Y,~Z\}$, and the gradient will be:

\begin{align}
\frac{\partial J( \bm{\theta} )}{\partial \theta _{i}} & =\bra{\psi _{i-1}}\frac{\partial U_{i}^{\dagger }( \theta _{i})}{\partial \theta _{i}} B_{i+1} U_{i}( \theta _{i})\ket{\psi _{i-1}}\notag \\
 & +\bra{\psi _{i-1}} U_{i}^{\dagger }( \theta _{i}) B_{i+1}\frac{\partial U_{i}( \theta _{i})}{\partial \theta _{i}}\ket{\psi _{i-1}}\notag \\
 & =\frac{i}{2}\left[\bra{\psi _{i-1}} U_{i}^{\dagger }( \theta _{i})( \hat{P}_{i} B_{i+1} -B_{i+1} \hat{P}_{i}) U_{i}( \theta _{i})\ket{\psi _{i-1}}\right]\notag \\
 & =\frac{i}{2}\left(\bra{\psi _{i-1}} U_{i}^{\dagger }( \theta _{i})[ \hat{P}_{i} ,B_{i+1}] U_{i}( \theta _{i})\ket{\psi _{i-1}}\right)
\end{align}\\

According to \cite{PhysRevA.98.032309},\\
\begin{align}
\frac{\partial J( \bm{\theta} )}{\partial \theta _{i}} & =\frac{1}{2}\left[\bra{\psi _{i-1}} U_{i}^{\dagger }( \theta _{i})\left( U_{i}^{\dagger }\left(\frac{\pi }{2}\right) B_{i+1} U_{i}\left(\frac{\pi }{2}\right) \right.\right. \notag \\& \left.\left. -U_{i}^{\dagger }\left( -\frac{\pi }{2}\right) B_{i+1} U_{i}\left( -\frac{\pi }{2}\right)\right) U_{i}( \theta _{i})\ket{\psi _{i-1}}\right]\notag \\
& =\frac{1}{2}\left[\bra{\psi _{i-1}} U_{i}^{\dagger }\left( \theta _{i} +\frac{\pi }{2}\right) B_{i+1} U_{i}\left( \theta _{i} +\frac{\pi }{2}\right)\ket{\psi _{i-1}} \right.\notag \\& \left. -\bra{\psi _{i-1}} U_{i}^{\dagger }\left( \theta _{i} -\frac{\pi }{2}\right) B_{i+1} U_{i}\left( \theta _{i} -\frac{\pi }{2}\right)\ket{\psi _{i-1}}\right]\notag \\
& =\frac{1}{2}\left[ J\left( \theta_{i} +\frac{\pi }{2}\right) -J\left( \theta_{i} -\frac{\pi }{2}\right)\right].
\end{align}\\

From the given evidence, we can infer that the gradient computed via both the adjoint differentiation and the parameter shift method coincides when no noise exists in the quantum circuit. These two methods employ distinct mathematical approaches to obtain the exact value of the gradient.
\end{appendices}

\bibliography{simple}

\end{document}